# Silver-Gold Bimetallic Alloy *versus* Core-Shell Nanoparticles: Implications for Plasmonic Enhancement and Photothermal Applications

Rituraj Borah,[a,b] Sammy W. Verbruggen[a,b]*

[a] Sustainable Energy, Air & Water Technology (DuEL), Department of Bioscience Engineering, University of Antwerp, Groenenborgerlaan 171, 2020 Antwerp, Belgium

[b] NANOlab Center of Excellence, University of Antwerp, Groenenborgerlaan 171, 2020 Antwerp, Belgium

*Sammy.Verbruggen@uantwerpen.be

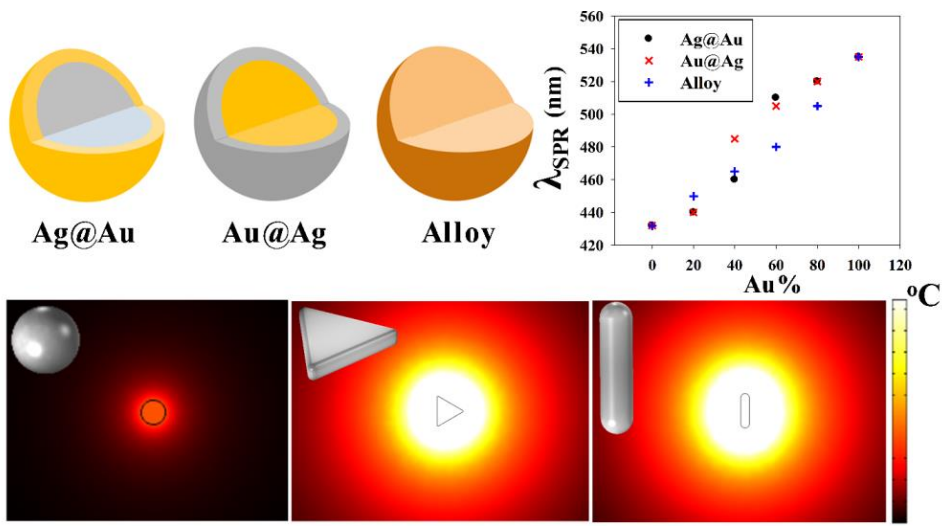

**Abstract**

Bimetallic plasmonic nanoparticles enable tuning of the optical response and chemical stability by variation of the composition. The present numerical simulation study compares Ag-Au alloy, Ag@Au core-shell, and Au@Ag core-shell bimetallic plasmonic nanoparticles of both spherical and anisotropic (nanotriangle and nanorods) shapes. By studying both spherical and anisotropic (with LSPR in the near-infrared region) shapes, cases with and without interband transitions of Au can be decoupled. Explicit comparisons are facilitated by numerical models supported by careful validation and examination of optical constants of Au-Ag alloys reported in literature. Although both Au-Ag core-shell and alloy nanoparticles exhibit an intermediary optical response between that of pure Ag and Au nanoparticles, there are noticeable differences in the spectral characteristics. Also, the effect of the bimetallic constitution in anisotropic nanoparticles is starkly different from that in spherical nanoparticles due to the absence of Au interband transitions in the former case. In general, the improved chemical stability of Ag nanoparticles by incorporation of Au comes with a cost of reduction in plasmonic enhancement, also applicable to anisotropic nanoparticles with a weaker effect. A photothermal heat transfer study confirms that increased absorption by the incorporation of Au in spherical Ag nanoparticles also results in an increased steady state temperature. On the other hand, anisotropic nanoparticles are inherently better absorbers, hence better photothermal sources and their photothermal properties are apparently not strongly affected by the incorporation of one metal in the other. This study of the optical/spectral and photothermal characteristics of bimetallic Au-Ag alloy *versus* core-shell nanoparticles provides a detailed physical insight for the development of new taylor-made plasmonic nanostructures.

**1. Introduction**

While the search for better plasmonic materials continues, most of plasmonic research so far has been based on Ag and Au nanostructures due to their excellent optical properties and ease of fabrication.[1] Both elements have their own benefits and shortcomings. The low interband transition energy of Au (2.3 eV) leads to energy loss by such transitions, whereas the larger energy gap of Ag (3.7 eV) does not allow this interband transition, thus implying more efficient use of incident light energy for hot electron generation.[2] Thus, while Ag suffers lower interband transition loss in the visible and near-infrared (NIR) range, Au is more resistant to oxidative degradation. For comparable sizes, Ag nanostructures have a Localized Surface Plasmon Resonance (LSPR) wavelength shorter than that of Au with stronger near field enhancement.[3, 4] Also, Ag nanoparticles have narrower and more symmetric plasmon absorption bands compared to Au.[5] While starkly different in optical response as pure metal nanoparticles, bimetallic Ag-Au nanoparticles show the possibility to tune the optical features tailored to specific applications by regulating the composition and structure.[6] Both core-shell and alloy nanoparticles of Ag-Au bimetallic systems have been shown to exhibit spectral characteristics that lay between those of pure Au and Ag nanoparticles.[7 - 10] Thus, bimetallic nanoparticles, constituted as both alloy and distinct core-shell phases, have been explored extensively, not just with Ag and Au, but also using other metals such as Cu, Pt and Pd, amongst others.[11] Considerable efforts have been devoted towards controlled synthesis of both core-shell and alloy nanoparticles with desired compositions. Generally, wet chemical synthesis of Au@Ag core-shell nanoparticles is straightforward and proceeds by the controlled reduction of an Ag precursor in the presence of Au nanoparticles, or seeds, that facilitate crystallization of Ag as a shell around the Au cores.[12] Conversely, Ag@Au core-shell nanoparticles are difficult to synthesize in a similar way. In the presence of Ag nanoparticles, $Au^{3+}$ ions trigger a galvanic replacement reaction that results in hollow structures.[13] It has been shown that by controlling the relative kinetics of the galvanic replacement and the reduction by a reducing agent, growth of Au shells on silver nanoparticles can be achieved with smooth morphology.[14-16] While a number of studies has reported direct synthesis methods for Ag-Au alloy nanoparticles by co-reduction of Ag and Au precursors,[8, 17] recent work shows that traditional co-reduction methods do not necessarily lead to fully homogeneous alloys.[18] Conversion of a hybrid Ag-Au nanostructure such as core-shell nanoparticles to alloys simply by heat treatment at temperatures as low as 250°C has also been shown to be a promising and facile synthesis route.[19-21] The intrinsic aspect that sets alloy nanoparticles apart from bimetallic core-shell nanoparticles is the overlapping of energy bands of the constituting elements in the alloy. Thus, the distinct plasmonic behavior of alloy nanostructures results from the altered dielectric properties of the mixed material.[22] In contrast, core-shell nanoparticles exhibit plasmonic behavior that is a hybrid of the individual plasmon modes of the core and the shell.[23]

This intrinsic difference and the ease of switching from core-shell to alloy naturally raises a question: Which is the better plasmonic material, alloy or core-shell? The answer clearly is subject to the intended application. In general, gold nanoparticles are preferred over silver nanoparticles for photothermal applications due to higher stability and absorption intensity of gold.[24, 25] Espinosa *et al*. recently showed that silver nanoparticles lose their photothermal properties with time due to chemical degradation inside human mesenchymal stem cells; but with a thin Au shell surrounding the Ag core, the nanoparticles can be stabilized completely.[24] On the other hand, spectral tunability of Ag-Au alloy nanoparticles enables wider coverage of the visible light spectrum for plasmon-assisted photocatalysis.[4] Similarly, the bimetallic constitution facilitates spectral tunability for SERS-based sensing.[26, 27] While spectral tunability is an advantage, plasmonic hot carrier generation is strongly dependent on the nanoparticle composition.[2]

In the present work, a comparative simulation study is performed on the plasmonic characteristics of Au-Ag bimetallic alloy and core-shell nanoparticles, and includes both spherical and anisotropic (nanorods and nanotriangles) shapes. The resulting photothermal effect of these nanoparticles is also evaluated and compared. In particular, absolute absorption/scattering spectra, near-fields and local temperature fields are studied to create a deeper physical insight into how the nanoparticle constitution can be perfected and optimally tuned to a given plasmonic application.

## 2. Problem Specification and Numerical Methodology

The present study involves electromagnetic simulations of isolated bimetallic spherical and anisotropic nanoparticles with a core-shell configuration as well as an alloy configuration of Ag and Au. Figure 1 shows schematics of spherical nanoparticles with corresponding Trasmission Electron Microscopy (TEM) images and Energy Dispersive X-Ray (EDX) analysis maps reproduced from literature. Since scattering is low for small nanoparticles (<50 nm), the size of the nanoparticles in this work was fixed at 60 nm, so that both absorption and scattering are significant over the entire range of studied compositions in order to draw meaningful comparisons. Similarly, the studied nanotriangles in this work are of 60 nm each side and 7 nm thickness, and the nanorods are 20 nm in diameter and 80 nm in length. The implications for smaller nanoparticles are discussed briefly in the later sections for the completeness of the results.

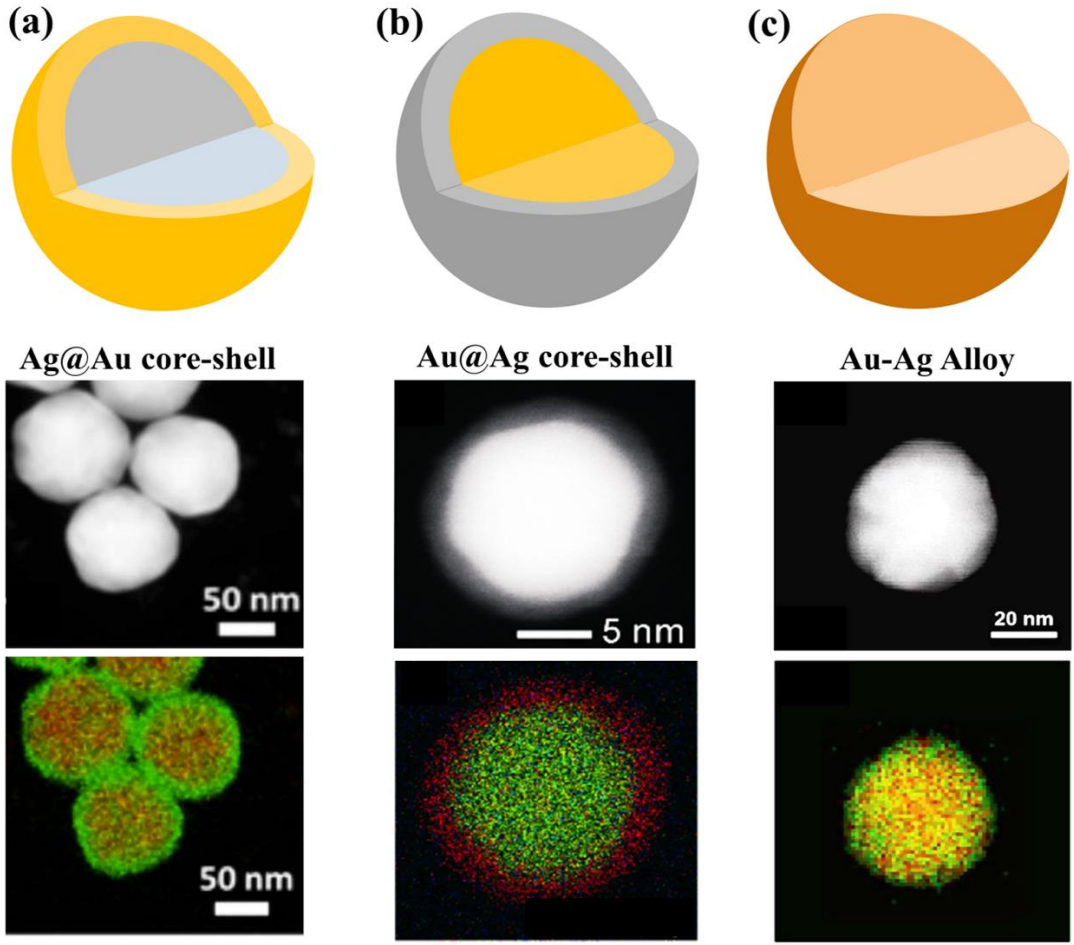

**Figure 1**. Schematic of spherical bi-metallic Au-Ag nanoparticles as (a) Ag@Au core-shell, (b) Au@Ag core-shell and (c) alloy with their TEM images and EDX mappings taken from references[15][12][17]. The images were adapted and reproduced with permission from American Chemical Society (© 2017), IOP Publishing (© 2012), Wiley-VCH (© 2019).

An FEM-based numerical framework COMSOL Multiphysics was implemented to solve the frequency domain form of Maxwell's equations:

$$\nabla \times (\mu_r^{-1} \nabla \times E_{sc}) - k_o^2 (\varepsilon_r - j\frac{\sigma}{\omega \varepsilon_o}) E_{sc} = 0 \qquad (1)$$

In equation (1), $\mu_r$, $\varepsilon_r$, and $\sigma$ denote material properties namely relative permeability, relative permittivity and electrical conductivity respectively. The scattered field solution of equation (1) is obtained in terms of the local scattered electric field, $E_{sc}$. The total field $E$ is the superposition of the scattered and incident electric fields, $E_{sc}$ and $E_{inc}$ respectively. A spherical computational domain was implemented with the nanoparticle at the center and a perfectly matched layer surrounding the domain. The perfectly matched layer was discretized by prismatic elements, while the computational domain including the nanoparticle was discretized by tetrahedral elements. While for plain spherical nanoparticles this discretization scheme has been proven to work well,[28] to ensure that it is also valid for the core-shell nanoparticles with a shell thickness as thin as 2.14 nm, a grid independence test was carried out. Comparing a 199043 element grid (minimum element size: 1.08 nm, maximum element size: 25 nm) with a 594120 element grid (minimum element size: 0.144 nm, maximum element size: 14.4 nm) in Figure S4 makes it clear that even distinctly different grid geometries provide excellent agreement in the spectra. Similar grid independence tests are also shown for a nanotriangle and a Ag@Au nanocube with 0.5 nm thin shell, Figure S4. This grid independence test validates the accuracy of the present numerical results.

The optical cross-sections were obtained from the scattered field solution. Absorption and scattering cross-sections for a nanoparticle can be defined as:

$$\sigma_{abs} = \frac{W_{abs}}{I} \tag{2}$$

$$\sigma_{sc} = \frac{W_{sc}}{I} \tag{3}$$

Where $W_{abs}$ and $W_{sc}$ are energy absorbed and scattered per unit time by the nanoparticle respectively, and $I$ is the intensity of incident light.

$$I = \frac{1}{2} c \varepsilon |E_{inc}|^2 \tag{4}$$

Where $\hat{k}$ denotes the direction of incident wave propagation. Now, the power absorbed and scattered in eq. (2) is obtained from the numerical solution as follows:

$$W_{abs} = \frac{1}{2} \iiint_V \mathrm{Re}[(\sigma E + j\omega D).E^* + j\omega B.H^*]dV \tag{5}$$

$$W_{sc} = \frac{1}{2} \oiint_S \mathrm{Re}[E_{sc} \times H_{sc}^*].ndS \tag{6}$$

Where, the superscript * and *D* stand for complex conjugate and displacement currents respectively. The integration in eq. (5) is throughout the nanoparticle volume and the surface integral in eq. (6) is over the nanoparticle surface. The built-in functions in COMSOL Multiphysics were used for these computations.

The heat transfer analysis for the evaluation of the photothermal effect of nanoparticles in water medium requires solving the steady state form of the heat equation:[29, 30]

$$\rho C_P\, U.\nabla T = k\nabla^2 T + q \qquad (7)$$

Here, *U* and *T* are the velocity vector and temperature, respectively. While, *k*, $\rho$, and $C_p$ denote material properties namely, thermal conductivity, density and specific heat capacity. In eq. (7), *q* is the rate of heat generation per unit volume due to absorption obtained as in eq. (5). For a given set values of these parameters, the problem is essentially simplified to the numerical solution of the conduction equation, as the velocity vector *U* in eq. (7) for the convective velocities of water can be neglected since convection will not be significant for an object of extremely small dimensions, also indicated by a very small Rayleigh number number (in the order of $10^{-14}$).[31] While, the convection does not influence the development of the temperature field, the investigation of the convective velocities, however small they are, is also important for applications in micr-environments. This aspect is not included in this work. A suitably large domain is chosen that imitates an infinite medium. In practice, the boundaries have to be extremely far as the temperature spatial decay scales as $1/r$, where *r* is the distance from the nanoparticle.[32] A spherical domain of size 2000 nm in diameter was found to be suitable as an infinite medium. The thermophysical properties were taken from relevant literature sources.[33, 34] Since the thermal conductivity of water is significantly low as compared to Au and Ag, the variability of the thermal conductivity of Au and Ag with temperature does not impact the conductive heat transfer characteristics. This is also implicit from the low value of the Biot number meaning the lump capacitance approximation is valid for both of the cases.

## 3. Results and Discussion

### 3.1 Model Validation

First, the present numerical model was validated using past experimental results. For alloy nanoparticles, the literature on reliable optical constant data is scent. Also, not all of the handful of such studies cover a wide spectral range or, more importantly, can reproduce experimental results accurately.[35, 36] Reliance on weighted average values as shown in some past literature is a too simplistic approach as the energy of the transitions changes with the composition.[8, 9, 37] In view of the literature available on this topic, Rioux *et al*. provides a rigorous model based on critical point analysis of the band structure of Au and Ag to show close agreement between theoretical and experimental values.[22] Their model overcomes the unphysicality of the simple Drude-CP (critical point) model reported by Etchegoin *et al*.[38, 39] inspite of the good quality of fit obtained. In this model the summation of Lorentzian oscillators representing individual (interband) transitions in Drude-Lorentz model has been replaced by an integration over of the joint density of states (jDOS), so that interband transition part to the dielectric function is a convolution of the jDOS with a simple Lorentzian oscillator whose strength is energy dependent. Thus, the shape of the dielectric function is determined by the shape of the jDOS which is depedent on the critical points *i.e*., Van Hove singularities in the jDOS. The two functions CP1 and CP2 defined by Rioux *et al*. for the inter-band transitions were obtained by the approximation of the joint density of states (jDOS). Upon defining this 10 parameter model of Ag and Au, the composition dependence of each patameter in the alloy has been modeled as a second-order polynomial giving rise to 3 sub-parameters. Thus, a 30 parameter model combined with the Drude free electron term was obtained for the composition depedent dielectric functions of Ag-Au alloy that overcomes the shortcomings of the previous literature. For the fitting of this model to experimental data, they implemented a genetic algoritm suited for multi-parameter optimization based on least square error.

Nonetheless, it is useful to compare the values reported in different studies, as presented in Figures S1, S2 and S3.[22, 40 - 42] For Ag-Au alloys, the agreement of Nishijima *et al*. and Peña-Rodríguez *et al*. with Rioux *et al*. is considerably better than Gong *et al*. over the visible light range, Figure S3. Further assessment of these data is done by comparing computed normalized scattering cross section spectra with experimental results reported by Patskovsky *et al*. in Figure 2(a).[43] Clearly, optical constants predicted by the theoretical model of Rioux *et al*. work well in estimating the extinction spectra of both pure as well as alloy nanoparticles. Au generally shows a better agreement than Ag using optical constants from different literature sources, as illustrated in Figures S5 and S6. When compared for 60 nm pure Ag nanoparticles, optical constants from Palik[44] appear to give large deviations from the rest, Figure S5. Also, the sharp quadrupolar peak in the absorption cross section is absent when optical constants from Palik were used. Apart from the experimental errors in the optical constants, the deviation of the computed spectral peak position from the experimental one, may also arise from the sensitivity to the surrounding environment, morphological irregularity of nanoparticles, instrumental errors and so on.[45] In Figure 2(b), computed extinction spectra using the dielectric constants of Rioux *et al*. are compared with experimental UV-Vis spectra of Ag nanocubes with a thin shell of Au reported by Yang *et al*.[46] The simulated results match the experimental data both quantitatively and qualitatively. With this overview of the available literature on the optical constants of Au and Ag alloy, it is reasonable to proceed with the model of Rioux *et al*. as it estimates the constants for any arbitrary composition and the agreement with experiments is shown to be the most satisfactory.

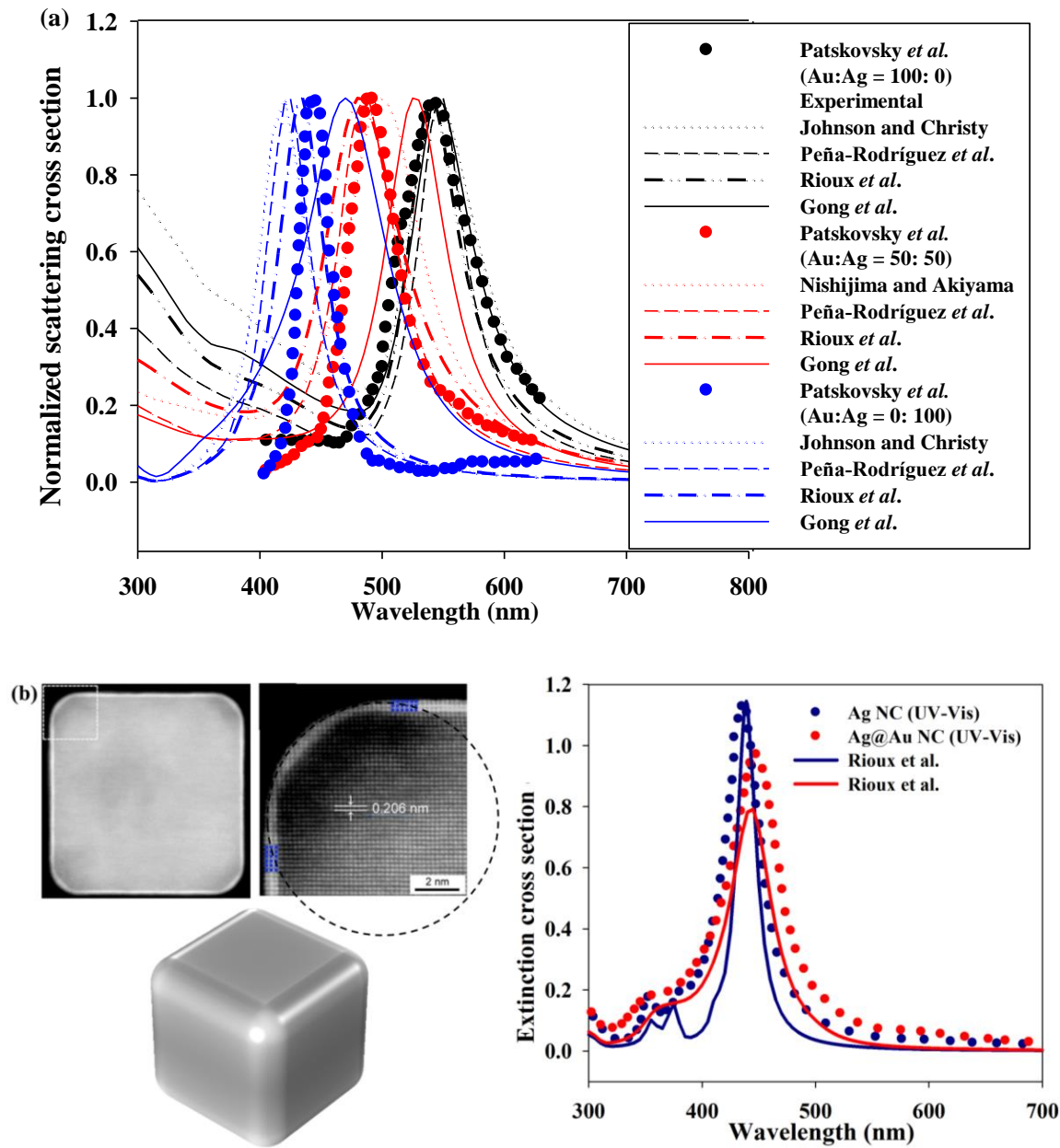

**Figure 2.** (a) Comparison of experimental scattering spectra of Patskovsky *et al.*[43] with computed spectra for optical constants from different literature sources: Johnson and Christy[47], Peña-Rodríguez *et al.*[41], Rioux *et al.*[22] and Gong *et al.*[42] for pure Au, pure Ag and 50:50 alloy Au:Ag (Note: Optical constants from Peña-Rodríguez *et al.* ae for 48:52 molar ratio of Au to Ag). Color codes: black for Au:Ag=100:0, red for Au:Ag=50:50, blue for Au:Ag=0:100 (b) TEM images of Ag nanocubes (AgNC) with thin Au protecting layer[46], visualization of 3D nanocube model in the present study, comparison of experimental extinction spectra with calculated spectra with constants from and Rioux *et al.* The TEM images are adapted and reproduced from American Chemical Society (© 2014).

### 3.2 Comparison of Plasmonic Properties: spherical nanoparticles

### 3.2.1 Absorption and scattering behavior

Before delving into bi-metallic nanoparticles, it is useful to analyze the role of the interband transitions and free electrons that eventually result in plasmonic excitation. As mentioned earlier, Au introduces such interband transitions, while Ag can be approximated as an entirely free electron contributor in the visible range. While the free elctrons result in plasmonic excitation, with consequent radiative and non-radiative damping, the interband transition contributes to "non-plasmonic" light capture and consequent thermalization. Discrimintating between both effects is possible by using a mathematical model for the wavelength dependent dielectric constants, where the parameters concerning individual effects can be selectively dropped. For this, the Drude-CP model that exploits critical point (CP) analysis[38, 39] has been shown to fit well with the dielectric constants of Au reported in Johnson and Christy[47], Figure 3(a) and (b). Figures 3(c) and (d) compare optical spectra obtained by separately considering the interband transitions (CP terms) and the free electrons (Drude term) in the dielectric constants with the spectrum that is obtained by considering the complete dielectric function. The position and the intensity of the LSPR is an outcome of these two effects coupled together. For a 60 nm nanosphere, the Drude free electron part in the absence of interband transitions yields plasmonic excitation that results in large radiative damping (scattering) and small non-radiative damping (absorption) at the LSPR. It is important to note that this purely plasmonic excitation is strongly dependent on nanoparticle's shape and size. In presence of the interband transition, however, the LSPR is red-shifted significantly, and the scattering component is drastically reduced with a proportionate enhancement in the absorption. Thus for 60 nm Au nanoparticles, the absorption cross section has a significant contribution from the interband transitions. However, whether it is the dominant contributor cannot be concluded directly. In contrast, the small size of a 20 nm nanoparticle facilitates strong non-radiative damping of the plasmons contributing to the absorption cross section far more significantly than interband transitions, Figure 3(e) and (f). Thus, while in the absence of interband transitions, the plasmonic excitation is remarkably strong as evidenced by the sharp absorption and scattering peaks; taking interband transition into account results in a drastic drop of the plasmonic excitation and the consequent optical cross sections. These competing effects of free electrons and interband transitions combined with the size and shape factors will also determine the optical response of Ag-Au bimetallic nanoparticles, as will be shown further.

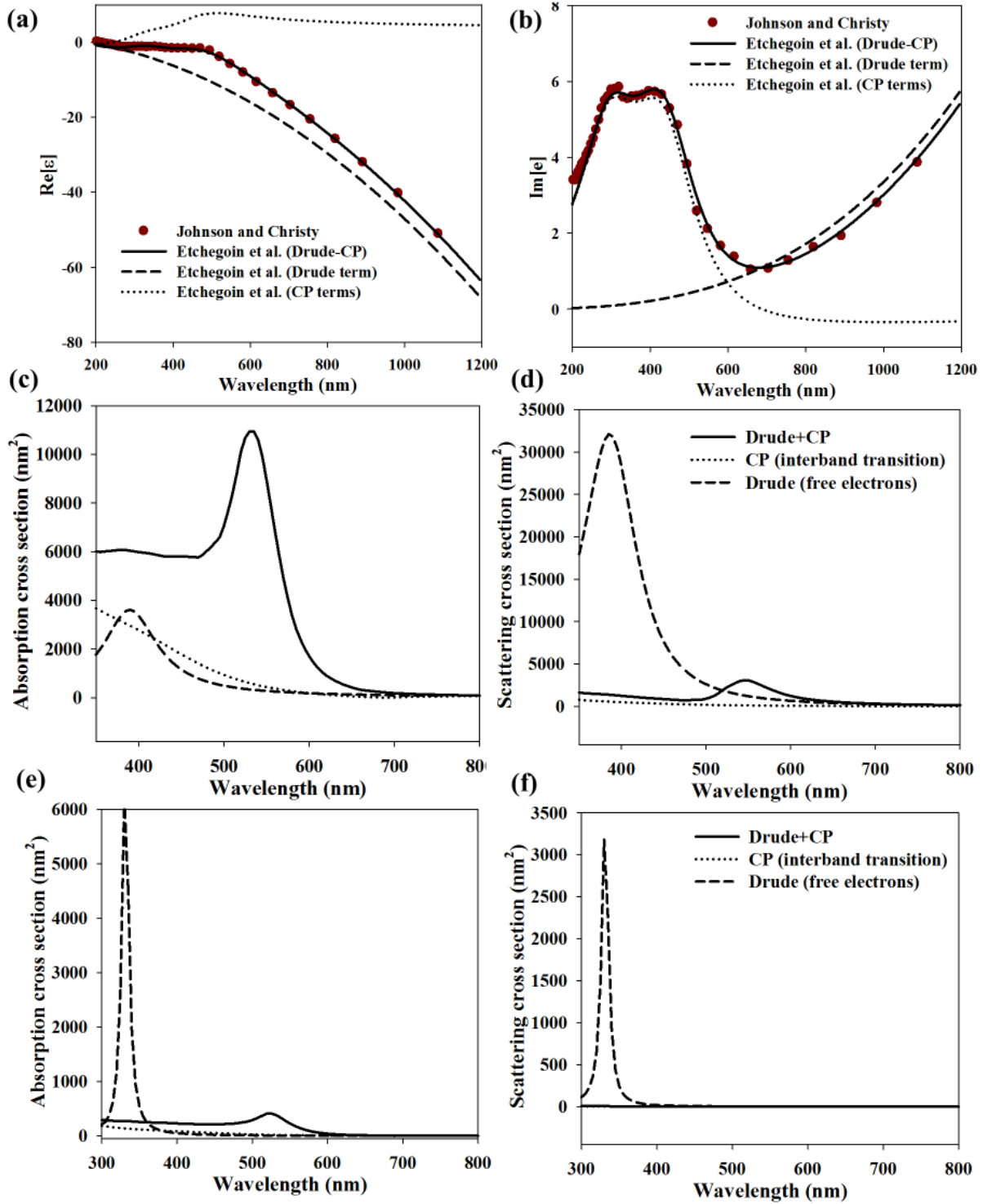

**Figure 3.** (a) Comparison of Johnsona and Christy[47] dielectric constants for Au with Drude-CP model as reported by Etchegoin *et al.*[38, 39]. The dielectric functions only considering the free electron Drude term and the interband transition CP term are also shown. (c, d) Comparison of absorption and scattering cross sections of a 60 nm Au nanoparticle with those calculated considering the Drude term and the CP terms alone. (e, f) Comparison of absorption and scattering cross sections of a 20 nm Au nanoparticle with those calculated considering the Drude term and the CP terms alone.

Figures 4 to 6 compare spherical Au-Ag alloy nanoparticles, Ag@Au core-shell nanoparticles, and Au@Ag core-shell nanoparticles, respectively, all of 60 nm in diameter and for different compositions. The general trends of the bimetallic constitution are discussed first before delving into the specific cases. In general, incorporation of Au in Ag nanoparticles induces a red shift of the plasmon band accompanied by a reduction in scattering, and increase in absorption as seen for both alloy and core-shell structures. From the discussion related to Figure 3, it can be attributed to the interband transitions in Au. Incorporation of even 20% Au significantly increases the absorption cross section for both alloy and core-shell nanoparticles accompanied by a corresponding drastic reduction in the scattering component. It is useful to note that the absorption component of the total energy extinction dissipates thermally while the scattered part is re-radiated to the surroundings. The absorption in Au is contributed by interband transitions from the *d*-band and plasmon decay, both resulting in hot carrier generation that eventually thermalize if not utilized in some other way.[48,49] Zheng *et al*. showed that hot carriers from plasmon decay are more energetic than those from interband transitions and are able to cross a ~1 eV Shottky potential barrier more efficiently. While in silver, interband transitions are absent.[50] The resulting absorption and scattering components are a commulative effect of these mechanisms and radiative damping of the plasmons. A 60 nm Ag nanoparticle is characterized by a strong radiative decay of plasmons (large scattering component) and weak thermalization. Incorporation of Au in Ag consequently introduces new interband transitions and reduction of the plasmonic excitation of the pure Ag state. The net result is an increase of the absorption component and a decrease in scattering. It is important here to bring up the case of 20 nm nanoparticles, for which Ag exhibits significantly higher absorption at LSPR than Au, as a consequence of strong decay or thermalization of plasmons and weak radiative damping limited by the small size. In this case, since the absorption is primarily from plasmon thermalization, incorporation of interband transitions via addition of another material would be detrimental to the absorption. Thus, incorporation of Au into (small) 20 nm Ag nanoparticles reduces the plasmonic excitation of the pure Ag state as Au introduces new interband transitions. The net result of this is reduced absorption. This in contrast to 60 nm Ag nanoparticles as discussed above, for which much of plasmonic excitation is lost to radiative decay in stead of thermalization and addition of even small amount Au only increases the absorption cross section by means of inter-band transitions. The resulting decrease in the overall extinction intensity upon introduction of Au in Ag is also clear from Figure S7. Importantly, this effect of interband transition is not seen in the case of anisotropic nanoparticles as will be discussed later. The change of optical intensities with the amount of Au incorporated does not seem to follow a continuously increasing trend. While for alloy nanoparticles, it can be attributed to the changes of the optical properties with different alloy compositions; for core-shell nanoparticles, it is rather attributed to the coupling of different sizes of the core and the shell.[23]

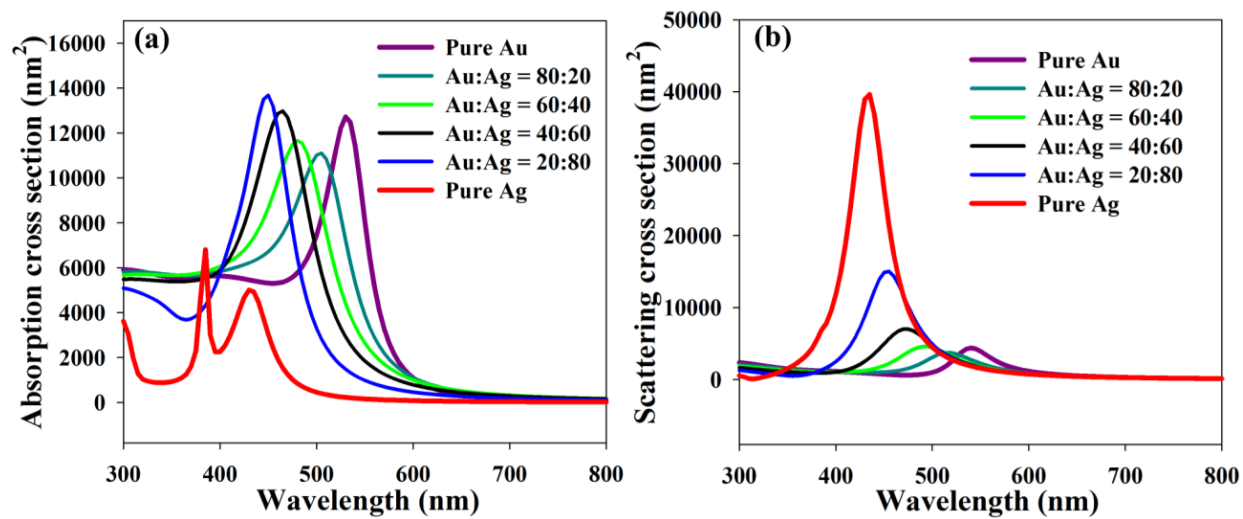

**Figure 4.** Absorption and scattering cross section of Ag-Au alloy nanoparticles with varying composition and fixed size of 60 nm.

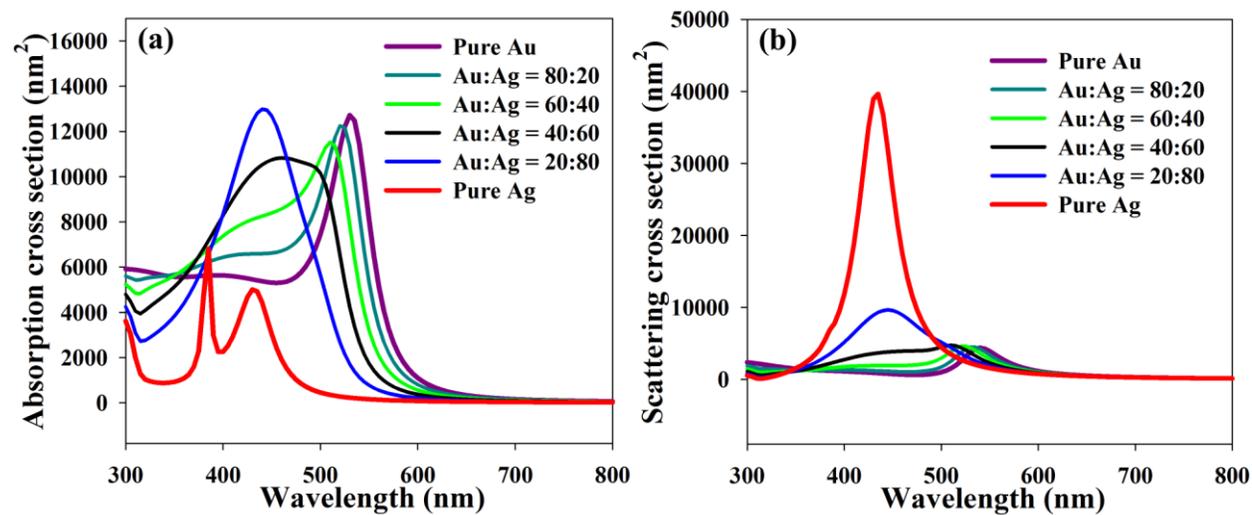

**Figure 5.** Absorption and scattering cross section of Ag@Au core-shell nanoparticles with varying composition and fixed size of 60 nm.

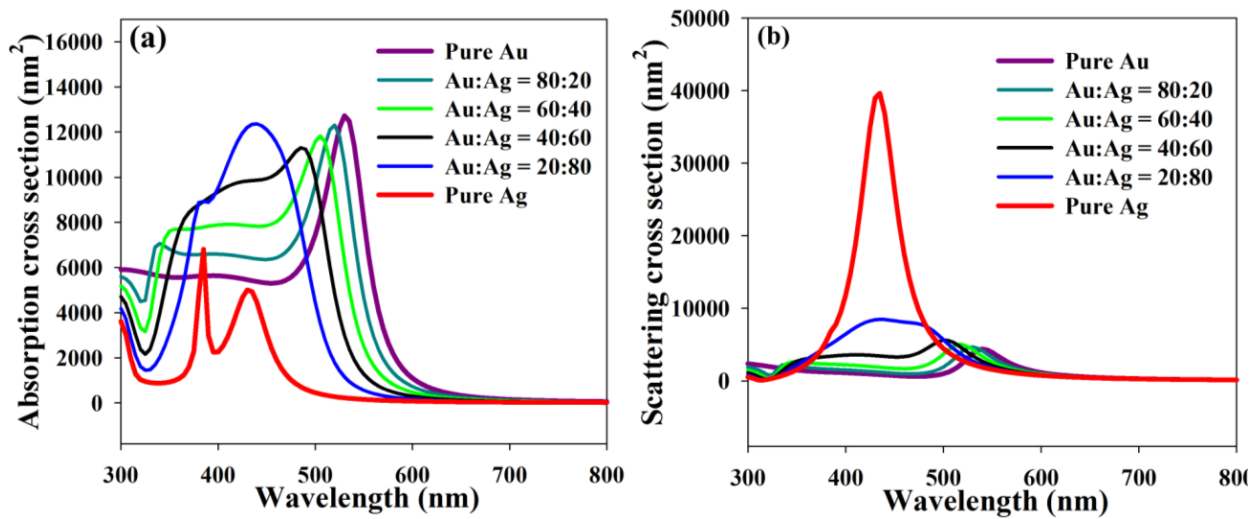

**Figure 6.** Absorption and scattering cross section of Au@Ag core-shell nanoparticles with varying composition and fixed size of 60 nm.

Further comparison of alloy with core-shell constitutions shows that the optical spectra of the alloy nanoparticles, Figure 4, are generally sharper (with one characteristic LSPR peak) than the core-shell ones, Figures 5 and 6. Experimental evidence of these trends has been reported in literature by means of UV-Vis data of bimetallic nanoparticles.[51, 52] The asymmetric broadening of the LSPR peaks of core-shell nanoparticles with different intermediate Ag:Au compositions, shown in Figure 5 and 6, is also in agreement with experimental observations for Au@Ag core-shell nanoparticles.[53 - 55] According to plasmon hybridization theory, the hybrid plasmon modes constituted by individual plasmon modes of core and shell are strongly dependent on the core and shell sizes. Thus, the optical spectra result from these hybrid modes. In contrast, an alloy nanoparticle has a homogeneous composition with its optical properties determined by the composition. Alloys exhibit optical properties that are intermediary between pure Ag and Au.[56, 57, 9, 10] It is noteworthy that recent research on Au-Ag alloy nanoparticles has shown that similar single dipolar LSPR peaks are also obtained for inhomogeneous bimetallic nanoparticles that attain a core-shell-like architecture.[17] These differences in the spectral characteristics between core-shell and alloy are well supported by experimental results.[53]

Considering the use of bimetallic nanoparticles for the tuning of the LSPR wavelength by variation of the composition, it is useful to understand the dependence of the spectral shift on the composition for both alloy and core-shell configurations. As apparent from Figure 7, the spectral shift upon introduction of Au in Ag is considerably different for all three configurations, Ag@Au, Au@Ag and alloy, taking into consideration the dipolar resonance peak. For alloy nanoparticles, the red-shift is proportionate to the amount of Au as alloy when moving from pure Ag to Au, also supported experimentally.[10, 8] For core-shell nanoparticles, the red-shift is not gradual. For instance, in Figure 7(a), going from 40% to 60% Au in the Ag@Au core-shell nanoparticle, the red shift of the absorption spectra is significantly larger than that while going from 60% to 80%. Also core-shell nanoparticles have irregular and wider spectra. For both absorption and scattering components in Figure 7(a) and (b), the LSPR peaks for the Au@Ag core-shell nanoparticle are located at larger wavelengths than for the alloy nanoparticles of the same overall composition.

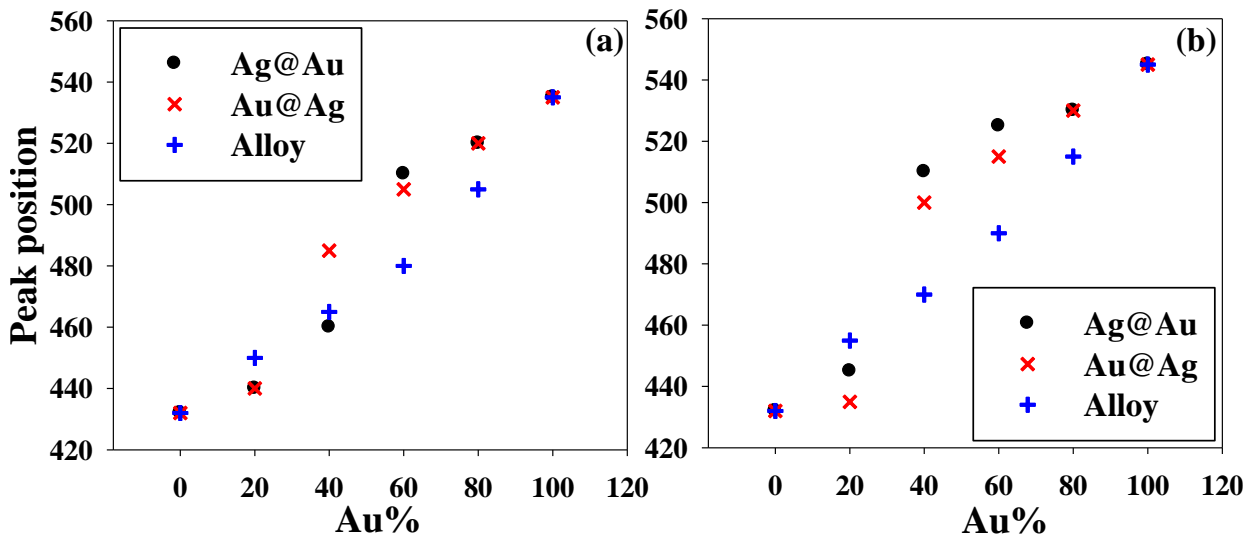

**Figure 7.** Spectral shift of absorption (a) and scattering (b) cross section with incorporation of Au for Ag@Au, Au@Ag and alloy nanoparticles of 60 nm in diameter.

**3.2.2 Near-field enhancement**

Despite the reduction of plasmon induced hot carriers upon introduction of Au in Ag, the chemical stabilization of Ag by Au is often quite useful. The result of energy consumption by interband transition is reduced plasmonic excitation. Zheng *et al*. quantified the hot carrier generation by plasmon decay separated from that by interband transitions and showed good correspondence with experiments.[48] Generally for spherical nanoparticles, the near-field enhancement is always weaker in gold nanospheres than in silver due to the contribution by interband transitions, Figure S8. Figure 8 shows that addition of 20% Au in Ag also significantly weakens the near-field enhancement around the thus-obtained alloy or core-shell nanoparticles when compared to the near-field enhancement of pure Ag nanoparticles. The alloy nanoparticle does not seem to be significantly different from the core-shell particle in terms of the attenuation of the near-field enhancement. Thus, stabilization of Ag nanoparticles in atmospheric environment by adding a thin Au shell has strong consequences with respect to the plasmonic properties.

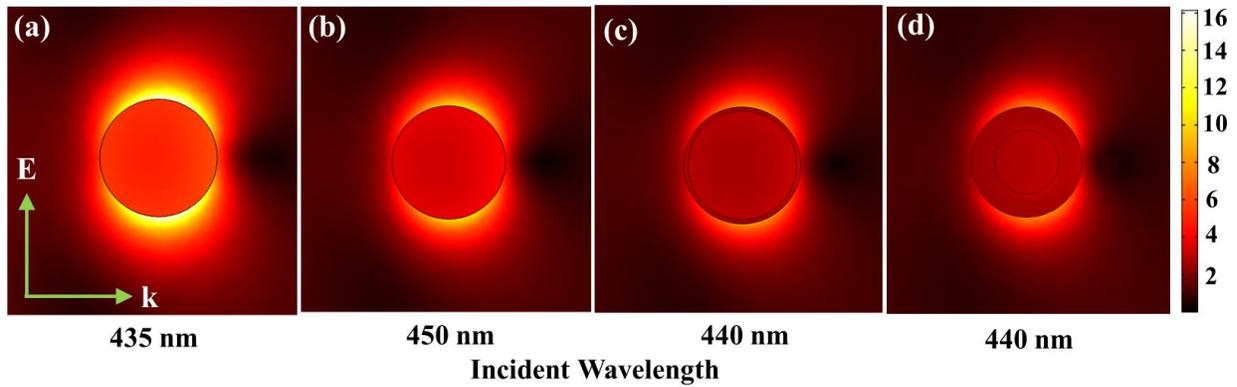

**Figure 8.** Near field enhancement with respect to the incident radiation around (a) pure Ag nanoparticle and (b) alloy (c) Ag@Au (d) Au@Ag nanoparticles of molar composition Au:Ag = 20:80 at their respective dipolar plasmon resonance modes.

Importantly, while 20% Au addition significantly alters both the spectral and near-field characteristics of a pure Ag nanoparticle, the thin shell of Au around Ag is only 2.14 nm in that case. Since the preceding comparisons were for a constant 60 nm nanoparticle size, incorporation of Au as the shell material entails a corresponding reduction of the Ag core size. To show that the drastic reduction in scattering and increase in absorption is not due to the shrinking of the core, optical spectra for a 60 nm Ag nanoparticle with a 2.14 nm Au shell around it (total diameter: 64.28 nm) were evaluated, Figure 9. Clearly, the effect of a slightly shrunk core is insignificant compared to that of the addition of the thin shell, resulting in strong changes in the optical spectra for a 2.14 nm thick shell surrounding a 60 nm particle core. The resulting optical response of the 64.28 nm nanoparticle differs from that of the 60 nm nanoparticle primarily due to the overall enlargement of the particle size.

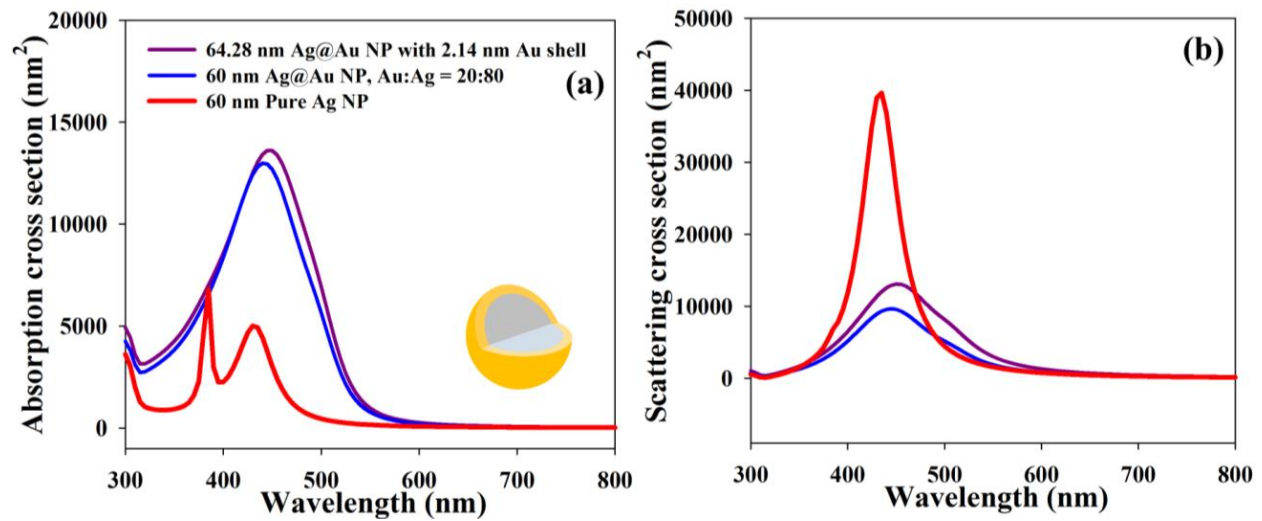

**Figure 9.** Effect of extra 2.14 nm thick Au shell on a 60 nm Ag nanoparticle compared to Au incorporation as shell (2.14) in a 60 nm nanoparticle for 20:80 molar composition of Au:Ag. (a) Absorption cross section (b) scattering cross section.

**3.3 Comparison of Optical Properties: anisotropic nanoparticles**

**3.3.1 Absorption and scattering behavior**

The discussion so far concerned nanospheres for which the LSPR wavelength falls in the wavelength range in which Au exhibits interband transitions. Thus, it is interesting to investigate the effect of bimetallic constitutions when the LSPR is in the near-infrared or infrared region where Au does not have interband transitions. In that regard, synthesis of anisotropic nanoparticles such as nanorods and nanotriangles, which exhibit LSPR in the near-infrared and infrared region, has seen considerable progress in the recent years.[58]

In continuation of the preceding discussion, the effect of incorporating 20% Au in an anisotropic Ag nanoparticle as a shell or alloy on the optical response is shown in Figure 10 and Figure 11. In this case the LSPR is located in the near-infrared and infrared region. The effect of addition of 20% Au in Ag in this case is interestingly different from the behavior that was observed in the case of nanospheres. In the absence of interband transitions, the absorption with the consequent thermalization is only caused by plasmon decay. Thus, apart from the obvious shift in plasmon band position, both the LSPR absorption and scattering intensities of an Ag nanorod, Figure 10, and a nanotriangle, Figure 11, are not significantly altered after adding an Au shell. Interestingly, when Au is added as alloy, the absorption is reduced a bit more, but the scattering is reduced dramatically. This rather strong change in the plasmonic response can be attributed to the altered band structure of the alloy. Experiments comparing the plasmonic hot carrier generation in Ag@Au core-shell and alloy nanorods or nanotriangles would thus provide useful insight into these observations. Regarding the spectral shift, while the Ag@Au core-shell nanorods and nanotriangles both exhibit a blue shift, in the case of alloying only the nanotriangles shows a blue shift but not the alloy nanorods. At this point it is also important to inspect the effect of corner sharpness changes even by addition of a thin layer, which for nanotriangles is obviously significant. Figure 12(a) and (b) show that layer by layer uniform enlargement of nanotriangles smoothens the corners leading to a blue shift of the spectra accompanied by an increase in both absorption and scattering. While for nanorods, adding uniform layers results in a usual red shift due to elongation and further remains unaffected by any edge effect, Figure 12(c, d). Thus, it is clear that the edge smoothening also has a significant role in the LSPR blue shift of nanotriangles when 20% Au is added as shell or alloy.

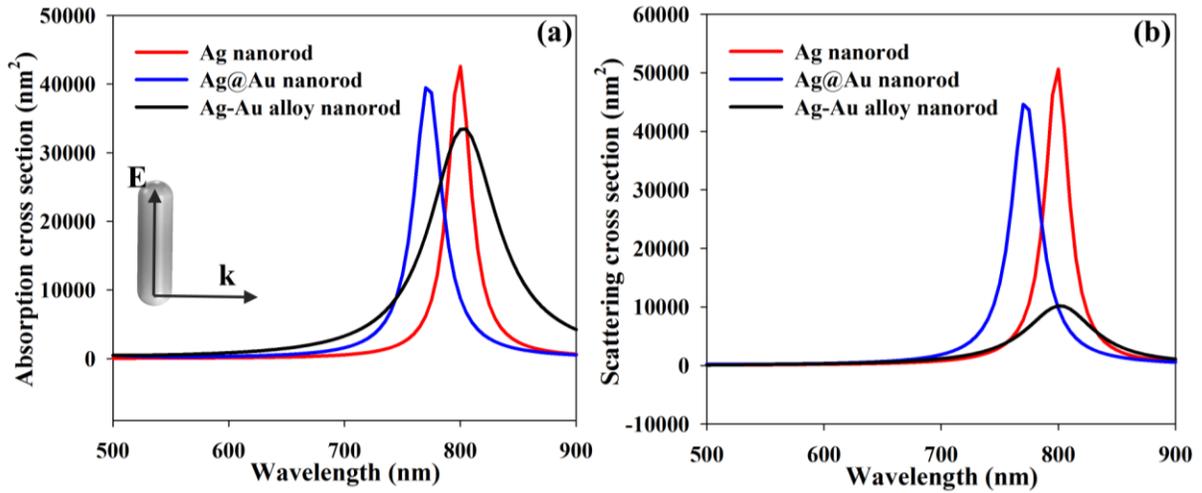

**Figure 10.** Effect of addition of 20% Au to an Ag nanorod (diameter: 20 nm, aspect ratio: 4) as a thin shell or alloy.

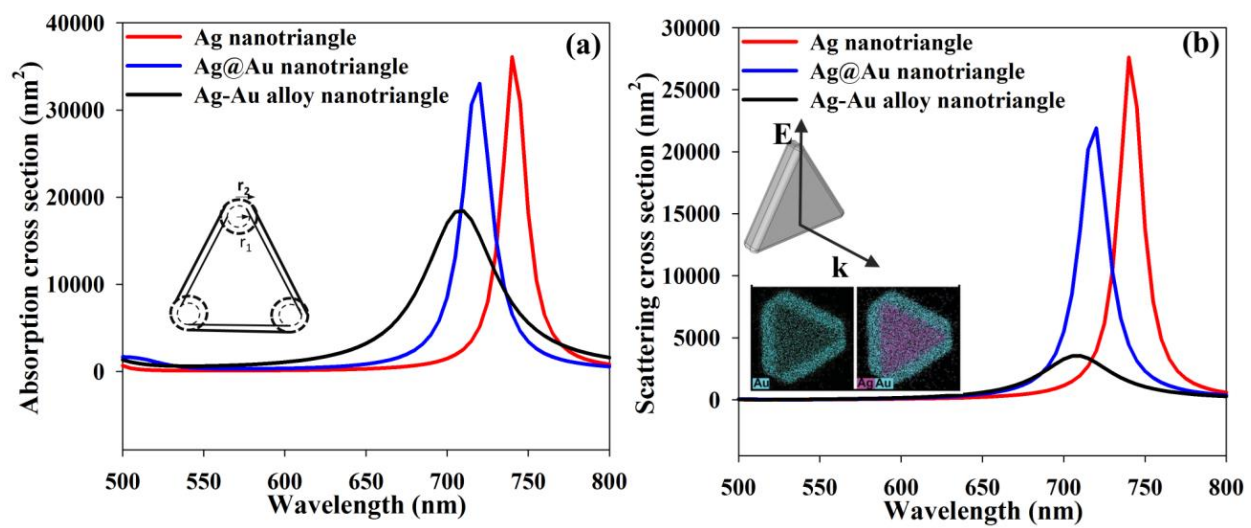

**Figure 11.** Effect of addition of 20% Au to an Ag nanotraingle (diameter: 20 nm, aspect ratio: 4) as a thin shell or alloy. Inset (a): schematic of a core-shell nanotriangle with corner curvatures $r_1$ and $r_2$ for the core and shell respectively. Inset (b): EDX mapping of synthesized Ag@Au core-shell nanotriangle reproduced from reference[16] with permission from Wiley-VCH (© 2019).

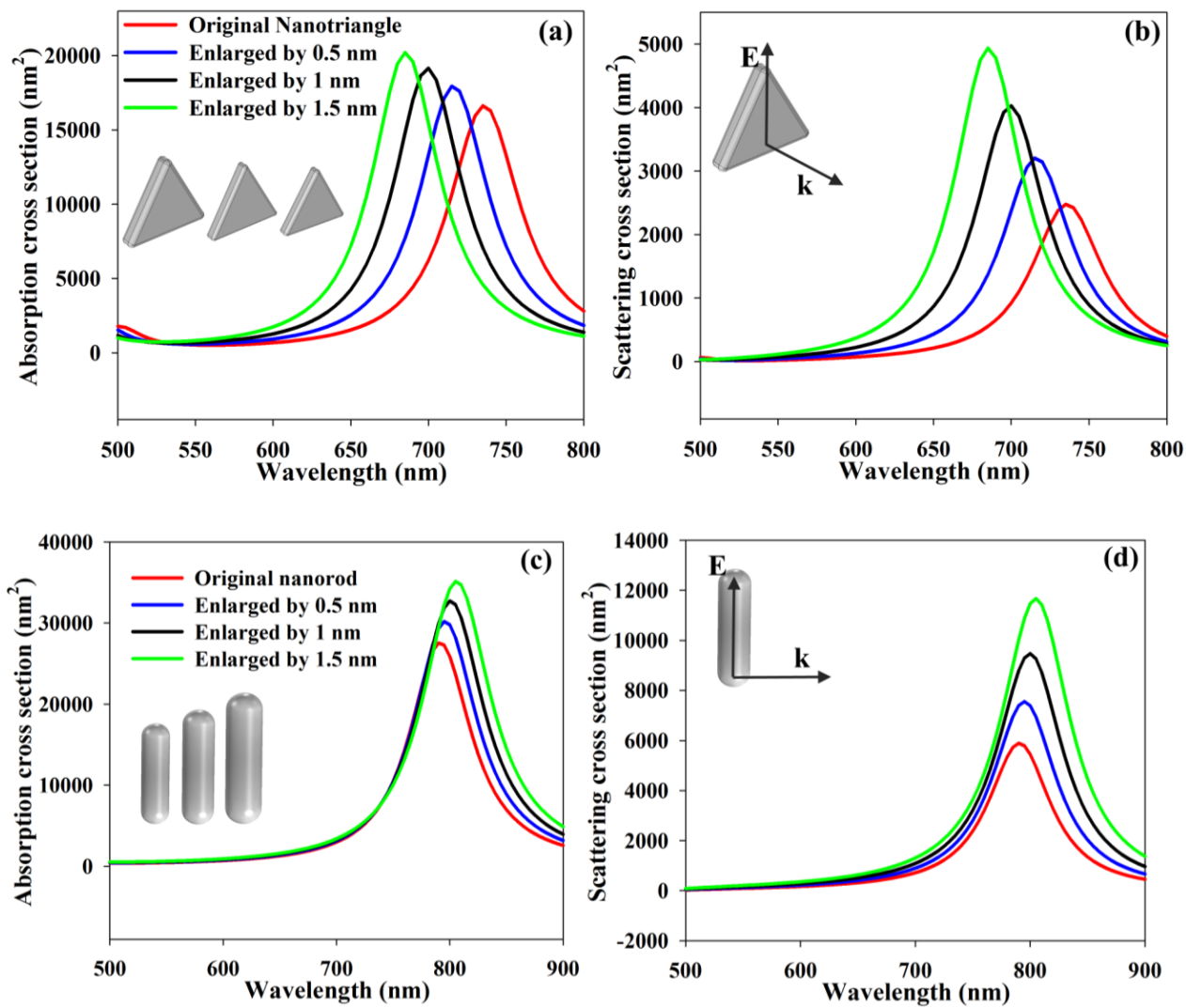

**Figure 12.** Effect of enlargement of Au:Ag = 20:80 alloy nanotriangle (a, b) and nanorod (c, d) by uniform addition of layers that reduces corner sharpness. Original Nanotriangle size: ~60 nm each side, 7 nm thick; original nanorod size: 20 nm in diameter, 80 nm in length.

It is also useful to look further into the effect of alloying of nanorods and nanotriangles for a fixed volume, hereby excluding the geometrical effects. Figure 13 and Figure 14 compare the optical spectra of alloy nanorods and nanotriangles, respectively, showing that Ag exhibits both stronger absorption and scattering than Au in the infrared region. Interestingly, absorption and scattering intensities of alloy nanoparticles are lower than for both pure Ag and Au nanoparticles. Although the LSPR is red shifted for Au, the shift has been found to be rather discontinuous when moving from pure Ag to Au.

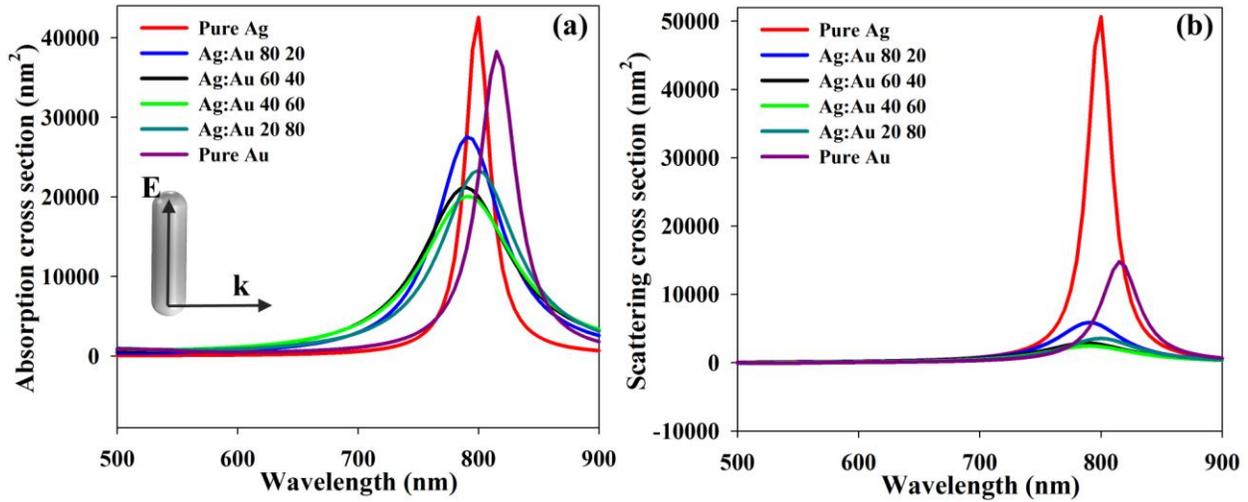

**Figure 13.** Absorption and scattering cross section of Ag-Au alloy nanorods with varying composition and fixed size of diameter 20 nm and aspect ratio 4.

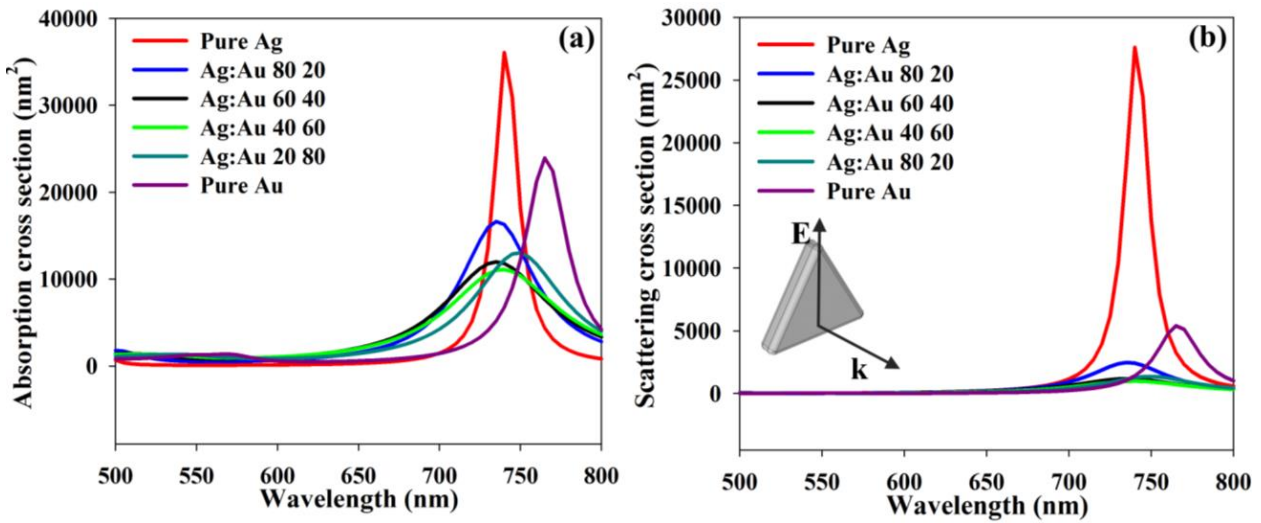

**Figure 14.** Absorption and scattering cross section of Ag-Au alloy nanotriangles with varying composition and fixed size of ~60 nm side length and 7 nm thickness.

The above analysis for anisotropic nanoparticles considered the direction of the incident electric field to be parallel with the nanorod's axis *i.e.*, longitudinal, while for nanotriangles the electric field is parallel to the triangular plane as shown in the Figures above. However, the alignment of the nanorods and nanotriangles with respect to the incident light has a significant effect on the plasmonic enhancement.[59, 60] While the plasmonic enhancement is dependent on the incident angle, the dependence is proportional. For instance, slight deviations from the longitudinal excitation for nanorods also results in slight reduction in the LSPR excitation.[60] Simularly, for nanotriangles, all the other orientations with respect to incident irradiation results in strong plasmonic extinction and the intensity is still more than half of that for the most favorable orientation.[59] Most importantly, the LSPR wavelength is independent of the orientation in all cases. Thus, the above results are still applicable in all different situations with varying degree of optical intensity. Also, certain applications of anisotropic nanoparticles particularly exploits this orientation dependence of plasmonic response.[60-62] It is useful to state that the problem of orientation dependence in certain applications can be solved with nanostars which exhibit strong LSPR at large wavelength irrespective of the orientation.[63]

### 3.3.2 Near-field enhancement

In the absence of interband transition of Au, it is also interesting to observe the near field characteristics of nanorods and nanotriangles in Figure 15. Clearly, due to edge effects, nanotriangles have significantly stronger near-field enhancement than nanorods. Overall, the optical cross sections in Figure 10 and 11 suggest that more "light" is captured by the nanorod. However, the sharper corners of the nanotriangles facilitates concentration of "light" over a smaller volume resulting in stronger near-field enhancement. So one must also keep in mind that uniform layer by layer enlargement of the nanorod will also strongly affect the near-field enhancement. It is clear from Figure 15 that the near-field enhancement is weakened by the addition of an Au shell to both the Ag nanorods and Ag nanotriangles, implying an overall reduction in plasmonic enhancement. Importantly, addition of Au as alloy weakens the near-field enhancement to a much greater extent. This along with the corresponding strong decrease in the scattering component (Figure 10 and 11) confirms a significant reduction in plasmonic excitation.

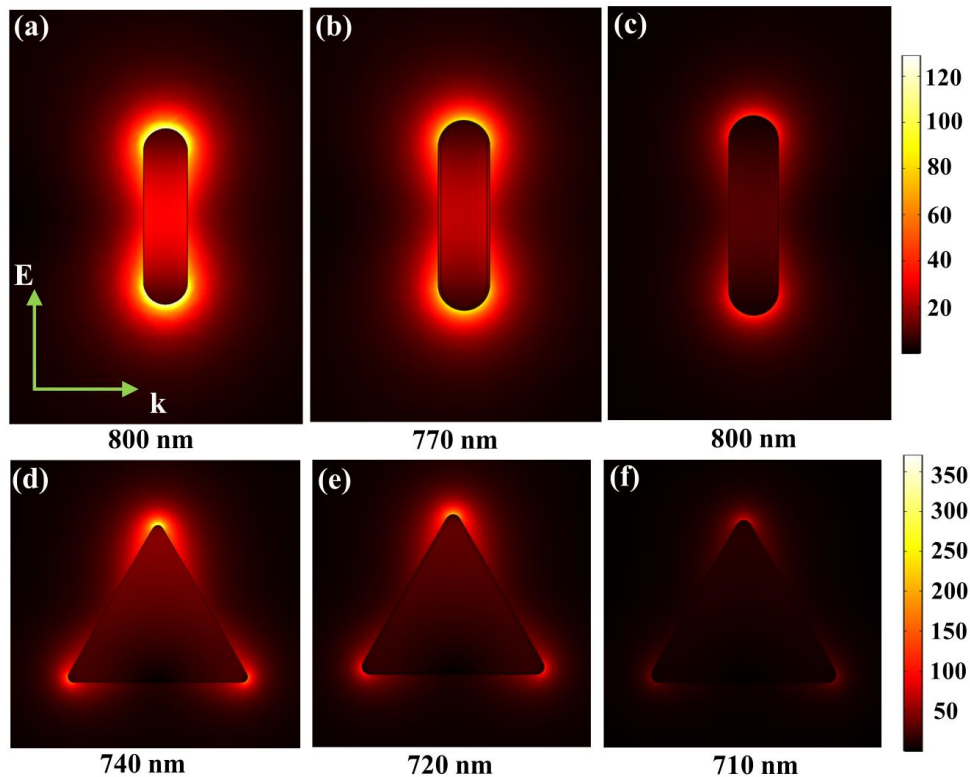

**Figure 15.** Near field enhancement with respect to the incident radiation around nanords of (a) pure Ag (b) Ag@Au (c) Alloy, and nanotriangles of (d) pure Ag (e) Ag@Au (f) Alloy. In correspondence to Figure 10 and 11, the Ag@Au core-shell (b,e) and alloy (c,f) nanoparticles are modifications of pure Ag nanoparticles in (a, d) by addition of 20% Au as a thin shell or as alloy. Nanotriangle size: ~60 nm each side, 7 nm thick; nanorod size: 20 nm in diameter, 80 nm in length.

### 3.4 Photothermal Effect

For spherical nanoparticles, while weakening of the optical response is generally disadvantageous for applications such as SERS, plasmonic phototcatalyis, *etc.*; the enhanced absorption and reduction of scattering due to Au incorporation is certainly advantageous for photothermal applications.[64] Likewise, the widening of the spectra in the case of core-shell nanoparticles provides a wider window of usable wavelengths. A large 60 nm Ag nanoparticle with high radiative loss and low absorption does not reach a maximum photothermal steady state temperature as high as a pure Au nanoparticle of the same size. In Figure 16, the steady state temperature of 45.5 °C of a 60 nm silver nanoparticle, Figure 16(a), is increased by ~70% to 77.2 °C simply by growing a thin gold shell of 2.14 nm around it, Figure 16(b). Also the LSPR band does not exhibit any significant shift, Figure 9. The steady-state temperature for a 60 nm Au nanoparticle under the same irradiation conditions is low (47.5 °C at 435 nm), unless the incident wavelength is changed to 530 nm in order to achieve a similar steady temperature 78 °C, Figure 16(c, d).

Such enhancement of the photothermal response of Ag nanoparticles by the incorporation of a thin shell of Au is an interesting aspect as the chemical stabilization of Ag nanoparticles by Au in cellular environments has already been demonstrated.[24] It is worth noting that the nanoparticle size is an important factor. For a smaller size of 20 nm, Ag exhibits strong non-radiative decay of plasmons that leads to considerably higher absorption than Au (for which the absorption also has contributions from inter-band transitions). In this case, the temperature of Ag will be considerably higher than that of Au. Also, incorporation of Au in Ag in this case will reduce the absorption of pure Ag. However, due to smaller absolute absorption for smaller nanoparticles, the steady state temperature is in turn significantly lower than for large nanoparticles.

The disadvantages of silver nanoparticles in the context of biomedical applications are the LSPR being in the high frequency region and small bandwidth (FWHM ~ 9 nm). Thus, LSPR tuning of silver nanostructures over a wide wavelength range is important for their utility in different applications. This is enabled by anisotropic nanoparticles such as nanoplates, nanotriangles, nanorods, *etc*.[58] For comparable sizes, Figure 17 shows that Ag nanotriangles and nanorods have significantly higher photothermal steady state temperature than spherical nanoparticles. Since the same color map scale of Figure 16 is used in Figure 17, one can see a much larger >75 °C region around the nanotriangle and nanorod. The implications of addition of Au as alloy or a protective layer around Ag nanorods and nanotriangles on the absorption are evident from Figure 10 and Figure 11. A thin protective layer of Au does not reduce the absorption greatly, implying a similar steady state temperature. On the other hand, addition as alloy results in a greater reduction in the absorption. However, the peak broadening provides a wider window of useful wavelengths.

Using the renewed physical insights from the present study, one may envisage the possibility of chemically stabilizing such anistropic Ag nanostructures with a thin gold shell.[24] In view of the present trends of bimetallic nanoparticle synthesis and photothermal applications of nanoparticles, the present results open up new avenues for further investigation.

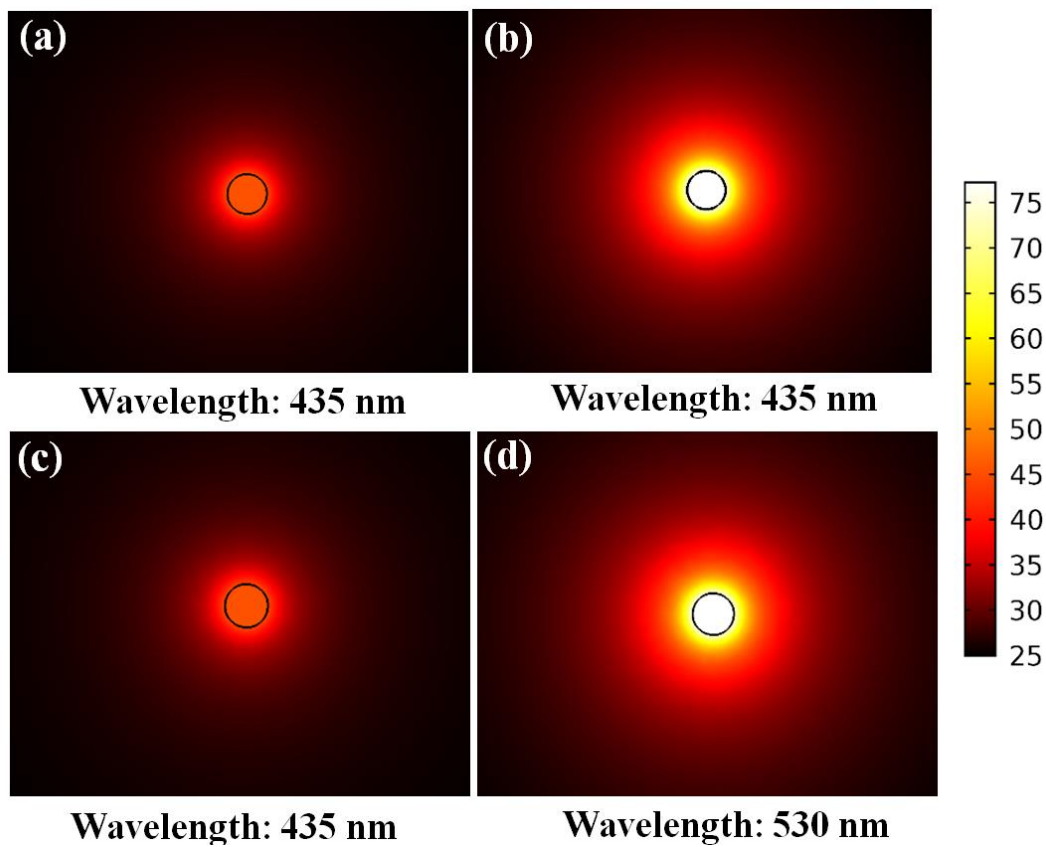

**Figure 16.** Steady state temperatures (°C) of (a) 60 nm Ag nanoparticle (b) 60 nm Ag nanoparticle with a 2.14 nm thin Au shell (c,d) 60 nm Au nanoparticle. The intensity of incident radiation is 1 mW/μm² for all the cases.

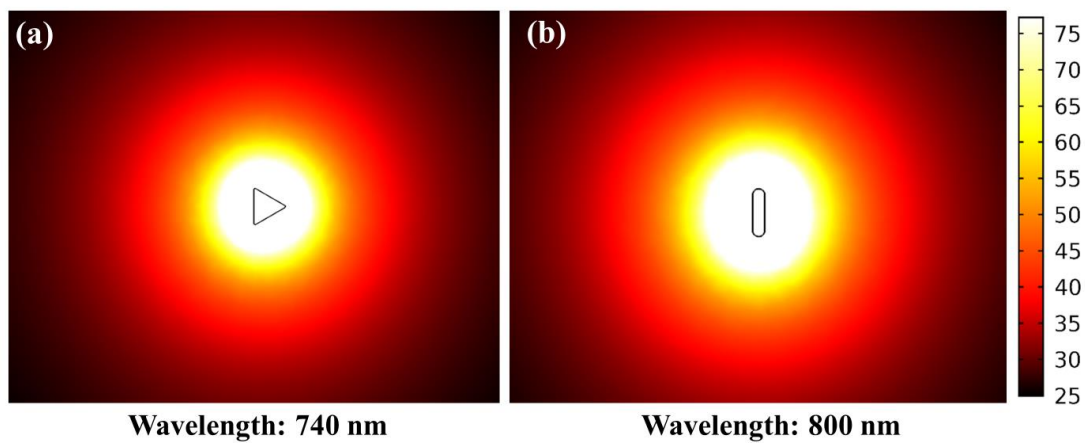

**Figure 17.** Steady state temperatures (°C) of an Ag nanotriangle (~60 nm each side, 7 nm thick) and Ag nanorod (20 nm in diameter, 80 nm in length). The intensity of incident radiation is 1 mW/μm² for all the cases. Note: the temperature scale is defined with respect to nanospheres in Figure 16 to make a visual comparison between the two.

## 4. Conclusion

The present study compares the plasmonic behavior of Ag-Au alloy and core-shell bimetallic nanoparticles and shows the implications on the optical response and the photothermal effect. The comparison includes spherical as well as anisotropic nanoparticles (rods and triangles) to cover structures with and without contribution of interband transitions of Au, respectively. The effects of bimetallic constitutions for these two cases are starkly different from each other.

For 60 nm spherical nanoparticles, the LSPR band shifts proportionally with changing composition in the alloy consitution; while for the core-shell constitution, the shift does not follow a proportional trend. Alloy nanoparticles display sharper spectral features with a characteristic LSPR peak in contrast to the widened spectra of core-shell nanoparticles. Incorporation of even a small amount of Au (20%) in Ag as alloy or core-shell significantly increases the absorption cross section due to interband transitions in Au. On the other hand, scattering is drastically reduced by the incorporation of the same amount of Au. Apart from the spectral characteristics, the optical intensities of alloy and core-shell nanoparticles are comparable. For nanorods and nanotriangles of comparable sizes, incorporation of Au in Ag results in a reduction of both absorption and scattering. While this reduction is not significant in a core-shell constitution, absorption and scattering are reduced dramatically in alloy nanoparticles. Especially, the drastic reduction in scattering suggests changes in the band structure by alloying that are detrimental to plasmonic enhancement. Also, unlike spherical particles, there is not a proportionate shift of the LSPR band with changing alloy composition from pure Ag to pure Au. Additionally for nanotriangles, the edge effects play a pivotal role in the LSPR intensity and spectral position. Thus, in anisotropic nanoparticles, geometrical effects become crucial while determining the plasmonic response. Increase in absorption and peak widening due to incorporation of Au in Ag improves the photothermal characteristics of 60 nm nanoparticles. In general, nanotriangles and nanorods are significantly better for photothermal applications due to much higher absorption in their optimal orientation. Addition of a thin Au shell around Ag nanorods or nanotriangles does not reduce the absorption intensities, thus the photothermal effect is well retained while the Ag structure is chemically stabilized. Thus, chemical protection of Ag nanoparticles by covering with an Au shell would be an efficient application strategy. These findings can serve as an input for the targeted synthesis of improved plasmonic nanomaterials. The numerical results and the obtained physical insights show interesting possibilities for the use and adaptation of bimetallic nanoparticles towards various plasmon-based applications.

**Supporting information**

Comparison of optical properties from various literature, grid-independence test, extinction spectra and near-enhancement of selected cases.

**Acknowledgements**

R.B. acknowledges financial support for the University of Antwerp Special Research Fund (BOF) for a DOCPRO4 doctoral scholarship.